\documentclass{article}
\usepackage{spconf,amsmath,graphicx}
\usepackage{mathptmx}
\usepackage{amsfonts}
\usepackage{hyperref}

\usepackage{enumitem}
\setlist{nosep, leftmargin=14pt}

\title{MultiViT2: A Data-augmented Multimodal Neuroimaging Prediction Framework via Latent Diffusion Model}
\name{Yuda Bi*, Sihan Jia*, Yutong Gao, Anees Abrol, Zening Fu and Vince D Calhoun\thanks{*These authors contributed equally.}}
\address{\textit{Tri-Institutional Center for Translational Research in Neuroimaging and Data Science (TReNDS)}}
\begin{document}

%\ninept
%
\maketitle
\begin{abstract}

Multimodal medical imaging integrates diverse data types, such as structural and functional neuroimaging, to provide complementary insights that enhance deep learning predictions and improve outcomes. This study focuses on a neuroimaging prediction framework based on both structural and functional neuroimaging data. We propose a next-generation prediction model, \textbf{MultiViT2}, which combines a pretrained representative learning base model with a vision transformer backbone for prediction output. Additionally, we developed a data augmentation module based on the latent diffusion model that enriches input data by generating augmented neuroimaging samples, thereby enhancing predictive performance through reduced overfitting and improved generalizability. We show that MultiViT2 significantly outperforms the first-generation model in schizophrenia classification accuracy and demonstrates strong scalability and portability.

\end{abstract}
\section{Introduction}
\label{sec:intro}

Multimodal data combining structural, functional, and genetic information has been increasingly used in medical imaging research for brain-related predictions and neurological disorder analysis, with advanced models like vision transformers (ViT) predicting diseases and identifying biomarkers \cite{dosovitskiy2020image, shaik2024multi, 10230385}. Despite progress, challenges remain in addressing data scarcity, integrating diverse data types, and developing portable foundation models. Schizophrenia, linked to structural, functional, and genomic factors, has been predicted using multimodal deep learning, but improving prediction accuracy and model portability continues to be critical for biomarker discovery and broader application to other brain disorders \cite{voineskos2024functional, di2024magnetic}.

The latent diffusion model (LDM) \cite{rombach2022high}, a framework based on diffusion models \cite{ho2020denoising}, excels in image generation and translation tasks. In unconditional setups, LDM can function as a data augmentation pipeline. While LDM has been widely used in natural image generation \cite{cao2024survey}, its application to 3D medical image generation, particularly brain MRI, remains underexplored \cite{kim2024adaptive, jiang2023cola}. Generating high-quality 3D MRI images requires addressing the inherent complexity and high dimensionality of the data \cite{wang2023inversesr}, making robust autoencoders crucial for capturing the intricate spatial relationships in 3D brain images.

Our current work extends previous multimodal models using structural MRI and functional network connectivity (FNC), offtering several key innovations and explorations. First, instead of relying solely on a ViT classification model, we integrate latent feature fusion module (LFFM), a portable pre-trained feature extraction module with a early state data fusion. We pre-trained the feature extraction module using the ABCD dataset (N=11,220), focusing on the 3D sMRI data. This pre-training significantly enhances the module's ability to represent complex 3D structures in sMRI, allowing for more accurate learning of features critical for neurological disease prediction. This approach also increases the adaptability of the model, allowing it to be effectively applied to different datasets and tasks.  In addition, we trained a 3D autoencoder with a KL divergence loss on the same ABCD dataset, addressing the limitations of existing LDM in the 3D MRI domain. This enhancement allows the model to better capture the inherent variability in 3D brain images and improves its overall generalization. Finally, we applied the LDM data augmentation model to the original dataset, which resulted in significant improvements in the accuracy and overall performance of the model, demonstrating the effectiveness of this augmentation technique in improving prediction results.

\begin{figure*}
    \centering
    \includegraphics[width=0.7\linewidth]{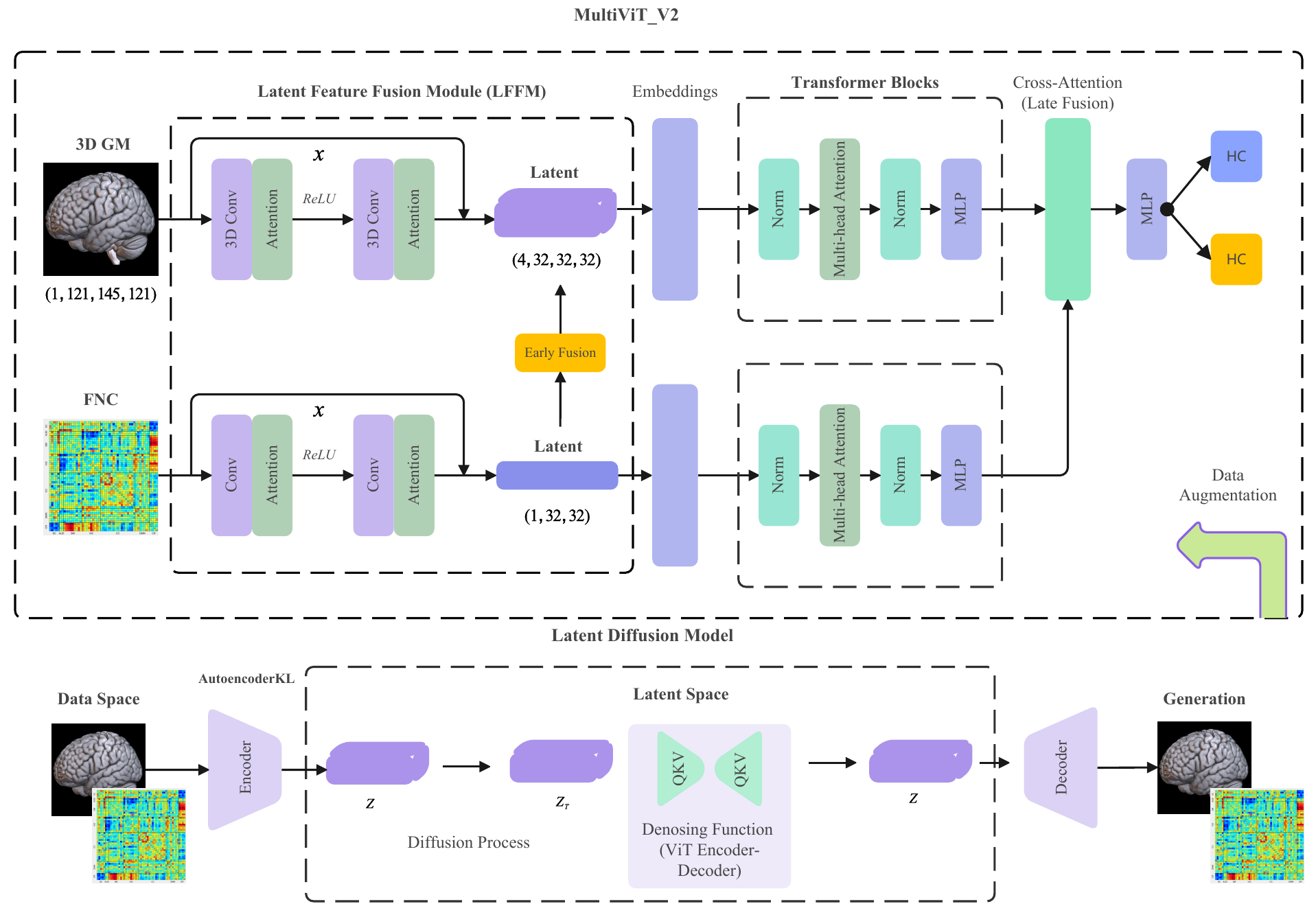}
    \caption{The overall pipeline of MultiViT2: We designed a multimodal hybrid model that combines a pretrained base model with a vision transformer backbone, effectively classifying structural and functional neuroimaging data while integrating a data augmentation module based on a latent diffusion model.}
    \label{fig:enter-label}
\end{figure*}

\section{Related Works}
\label{sec:rw}

Deep learning models for multimodal fusion have seen significant advancements in the field of brain disease research. Qiu et al. \cite{qiu20243d} introduces a 3D multimodal fusion network named MDL-Net, designed for the early diagnosis of Alzheimer'’s Disease (AD). MDL-Net effectively integrates 3D multimodal imaging data, including structural MRI and PET, to construct a deep learning model with richer features for AD diagnosis and brain region analysis. Alrawis et al. \cite{alrawis2024bridging} proposes a multimodal approach that integrates EEG and MRI data, leveraging their complementarity to enhance diagnostic accuracy for early Parkinson's disease (PD) diagnosis, and outperforming traditional single-modal and multimodal methods. Zhang et al. \cite{zhang2024pyramid} introduces PA-Net, a generative adversarial network with a pyramid attention mechanism, to address missing PET data in Alzheimer's disease classification. By generating realistic PET images and integrating MRI gray matter with PET metabolic information, the method reduces network input parameters and improves classification accuracy.

\section{Methods}

\subsection{Latent Diffusion Model with Autoencoder}

\begin{table*}[h]
\centering
\caption{Model performances for baselines, abligations and MultiViT2}
\begin{tabular}{llllllll}
\hline
\textbf{Name}      & \textbf{Main Model} & \textbf{Data} & \textbf{LFFM} & \textbf{Augmented} & \textbf{Accuracy} & \textbf{AUC}  \\ \hline
Baseline1   & ViT             & MRI        & -          & -          & 0.783       & 0.784             \\ 
Baseline2   & MultiViT\cite{10230385}             & FNC        & No          & No          & 0.831       & 0.833             \\ 
Abligation1 & CNN/ViT         & MRI/FNC    & No         & Yes        & 0.854       & 0.856              \\
Abligation2 & CNN/ViT         & MRI/FNC    & Yes        & No         & 0.853       & 0.854              \\  
MultiViT2   & MultiViT2         & MRI/FNC    & Yes        & Yes        & 0.866       & 0.866              \\ \hline
\end{tabular}
\end{table*}

The latent diffusion model integrates an autoencoder to compress high-dimensional data \(\mathbf{x} \in \mathbb{R}^D\) into a lower-dimensional latent representation \(\mathbf{z} \in \mathbb{R}^d\), where \(d \ll D\). The autoencoder consists of two parts: an encoder (\(E\)) that maps the input data to the latent space, and a decoder (\(D\)) that reconstructs the data from the latent representation. Specifically, the encoder performs the mapping \(E: \mathbb{R}^D \rightarrow \mathbb{R}^d\), resulting in \(\mathbf{z} = E(\mathbf{x})\), while the decoder maps the latent variables back to the original data space \(\hat{\mathbf{x}} = D(\mathbf{z})\).

In the latent space, a diffusion process is applied, where Gaussian noise is progressively added to the latent variables over \(T\) timesteps. This forward diffusion is modeled as a Markov chain, defined by:

\[
q(\mathbf{z}_t \mid \mathbf{z}_{t-1}) = \mathcal{N}(\mathbf{z}_t; \sqrt{1 - \beta_t} \mathbf{z}_{t-1}, \beta_t \mathbf{I}),
\]

where \(\beta_t\) is the variance schedule and \(\mathcal{N}\) denotes the normal distribution. The overall forward process over all timesteps can be written as:

\[
q(\mathbf{z}_{1:T} \mid \mathbf{z}_0) = \prod_{t=1}^T q(\mathbf{z}_t \mid \mathbf{z}_{t-1}).
\]

After the forward process, the reverse diffusion process is used to recover the original latent variable \(\mathbf{z}_0\) from the noisy latent variable \(\mathbf{z}_T\). This reverse process is parameterized by neural networks, following:

\[
p_\theta(\mathbf{z}_{t-1} \mid \mathbf{z}_t) = \mathcal{N}(\mathbf{z}_{t-1}; \mu_\theta(\mathbf{z}_t, t), \Sigma_\theta(\mathbf{z}_t, t)),
\]

where \(\mu_\theta\) and \(\Sigma_\theta\) are learned functions. The reverse diffusion process starts from \(\mathbf{z}_T \sim \mathcal{N}(0, \mathbf{I})\) and iteratively denoises the latent variable to reconstruct \(\mathbf{z}_0\). Finally, the decoder from the autoencoder is used to transform the denoised latent variable back to the original data space, yielding \(\hat{\mathbf{x}} = D(\mathbf{z}_0)\).

The training of the autoencoder is critical for latent diffusion model, which involves multiple loss functions. The \textbf{Reconstruction Loss} (\(\mathcal{L}_{\text{recon}}\)) ensures the accurate reconstruction of the input data:

\[
\mathcal{L}_{\text{recon}} = \mathbb{E}_{\mathbf{x}} \left[ \| \mathbf{x} - \hat{\mathbf{x}} \|_2^2 \right].
\]

In the case of variational autoencoders (VAEs), the \textbf{Kullback-Leibler Divergence Loss} (\(\mathcal{L}_{\text{KL}}\)) regularizes the latent space to align with a prior distribution:

\[
\mathcal{L}_{\text{KL}} = \frac{1}{2} \sum_{i=1}^d \left( \mu_i^2 + \sigma_i^2 - \log \sigma_i^2 - 1 \right),
\]
where \(\mu_i\) and \(\sigma_i\) are the mean and standard deviation of the latent variables.

The overall training objective combines these loss functions, with \(\lambda_{\text{recon}}\) and \(\lambda_{\text{KL}}\) controlling their relative contributions:

\[
\mathcal{L}_{\text{total}} = \lambda_{\text{recon}} \mathcal{L}_{\text{recon}} + \lambda_{\text{KL}} \mathcal{L}_{\text{KL}}.
\] 

The latent diffusion model leverages the compression and reconstruction capabilities of the autoencoder, and the diffusion process operates within the compressed latent space, offering a more efficient and structured generative process.

\subsection{MultiViT2 Architecture}

The MultiViT2 architecture is designed with a latent feature fusion module (LFFM) and a ViT classification pipeline, enhanced by late fusion through cross-attention mechanisms.

\textbf{Latent Feature Fusion Module (LFFM)}: The architecture begins with a 3D input tensor \(\mathbf{X} \in \mathbb{R}^{L \times W \times H}\), which is processed by the Latent Feature Fusion Module (LFFM). In this module, the high-dimensional input is transformed into a latent representation using a pretrained feature extraction network. Subsequently, FNC information is integrated via early fusion of the latent representations. Specifically, the sMRI latent tensor \(\mathbf{Z}\) is reduced along one spatial dimension to derive the FNC latent tensor \(\mathbf{Z}'\). A convolutional layer fuses \(\mathbf{Z}'\) with \(\mathbf{Z}\), aligning their spatial dimensions and enabling the model to learn meaningful interactions between the two modalities:

\[
\mathbf{Z}_{\text{fused}} = \text{ConvFusion}(\mathbf{Z}, \mathbf{Z}'),
\]

where \(\text{ConvFusion}(\mathbf{Z}, \mathbf{Z}')\) denotes the convolutional operation that produces the fused latent representation \(\mathbf{Z}_{\text{fused}}\).

\textbf{ViT Classification Pipeline}: The fused latent representation \(\mathbf{Z}_{\text{fused}}\) is tokenized and passed through a series of transformer blocks to capture higher-order features. Similarly, the FNC latent tensor \(\mathbf{Z}'\) is processed through transformer blocks to obtain an enhanced representation \(\mathbf{Z}_{\text{FNC}}\). Afterward, a cross-attention mechanism is applied to integrate complementary information from both \(\mathbf{Z}_{\text{fused}}\) and \(\mathbf{Z}_{\text{FNC}}\):

\[
\mathbf{Z}_{\text{Final}} = \text{CrossAttention}(\text{T}(\mathbf{Z}_{\text{fused}}), \text{T}(\mathbf{Z}')),
\]

where \(\text{T}(\cdot)\) represents the transformer block operations. The final latent representation, \(\mathbf{Z}_{\text{Final}}\), is then used for classification:

\[
\mathbf{Y} = \text{softmax}(\text{MLP}(\mathbf{Z}_{\text{Final}})),
\]

where \(\mathbf{Y}\) represents the predicted class probabilities. This pipeline effectively combines information from multiple modalities, ensuring robust classification performance.

\section{Experiments}
\label{sec:exp}
\subsection{Dataset and Pre-processing}

We utilized the \href{https://abcdstudy.org/}{ABCD dataset} for our experiments. First, T1-weighted MRI data was segmented using SPM12 to extract gray matter regions. Next, group ICA was applied to the fMRI data to obtain the FNC matrix. The ABCD dataset, including gray matter and FNC information, was used to train both the autoencoder and the LFFM, allowing the model to effectively learn representations from both structural and functional neuroimaging data. For the downstream task of schizophrenia prediction with MultiViT2, we employed two comprehensive schizophrenia-related datasets. The combined dataset included data from three international studies (fBIRN, MPRC, and COBRE) and several hospitals in China. In total, the dataset consisted of 1,642 participants: 803 healthy controls and 839 individuals diagnosed with schizophrenia. Resting-state fMRI (rsfMRI) data were acquired using 3.0 Tesla scanners across multiple sites, with standard echo-planar imaging (EPI) sequences (TR/TE approximately 2000/30 ms, voxel sizes ranging from 3 × 3 × 3 mm to 3.75 × 3.75 × 4.5 mm).

\subsection{Experimental Setup}

We primarily conducted comparison and ablation studies to evaluate the performance of our proposed approach. We created two baselines: Baseline 1 used unimodal models based on sMRI and FNC data separately. Baseline 2 used MultiViT1 \cite{10230385}, a basic multimodal model that lacked all of the innovations introduced in the second-generation model. In the comparison experiments, we demonstrated that the latest MultiViT2 architecture outperformed both Baseline 1 and Baseline 2, even when the pretrained model and data augmentation components were removed. However, the ablation studies were more critical to this research. We conducted two ablation experiments to assess the contributions of different components of our model. Ablation 1: We removed the LFFM but keep the LDM-based data augmentation process. Ablation 2: We kept the LFFM and ViT classifier but removed the LDM-based data augmentation component. These ablation experiments allowed us to analyze the individual contributions of the pre-trained module and data augmentation to the overall model performance.

\subsection{Training, Evaluation and Visualization}

To train the baseline models, we employed the AdamW optimizer and a ReduceLROnPlateau scheduler with a learning rate of 3e-4, training for a total of 150 epochs. For the MultiViT2 and its ablation models, we similarly used the AdamW optimizer but incorporated a 20-epoch warm-up phase alongside the ReduceLROnPlateau mechanism. The learning rate was maintained at 3e-4, with models also trained for 150 epochs. During the evaluation, we conducted 5-fold cross-validation, recording both accuracy and AUC metrics for each fold to assess model performance.

To visualize the importance maps generated by the attention mechanism on 3D sMRI data, we applied an attention-weighting method. These highlighted regions likely correspond to the model’s response to the integration of functional data, revealing the regions of interest (ROIs) in the structural data most strongly associated with schizophrenia. Additionally, we averaged attention weights across all transformer encoder layers and self-attention heads, providing a more comprehensive representation of the attention mechanism's effects throughout the model.

\section{Results}
\label{sec:results}

\subsection{Basic Experimental Results}

The results of our experiments show that the MultiViT2 model achieved superior performance compared to both the baseline and ablation models. As shown in Table 1, MultiViT2 outperformed Baseline 1 and Baseline 2 with improvements in both accuracy and AUC metrics, reaching 0.866 for both measures. In the ablation studies, removal of either the LFFM (Abligation1) or the data augmentation component (Abligation2) resulted in slightly reduced performance, indicating that both components contribute significantly to the effectiveness of the model. Abligation1 had an accuracy of 0.854, while Abligation2 had an accuracy of 0.853, both lower than MultiViT2. These results confirm that the integration of LFFM and LDM-based data augmentation improves the model's ability to effectively classify multimodal neuroimaging data.

\subsection{Saliency Map Visualizations}

Our structural brain saliency maps highlight the brain regions that contribute most to the model's classification of schizophrenia. According to our analysis, regions such as the cerebellum, caudate, precuneus, and superior frontal orbital gyrus show strong relevance to schizophrenia prediction. These areas are commonly associated with motor control, cognitive functions, self-awareness, and emotional regulation, which are often impaired in people with schizophrenia. For example, the cerebellum is involved in motor coordination and cognitive processing, while the caudate plays a key role in movement regulation and learning. The precuneus is critical for self-awareness and spatial processing, and the superior frontal orbital gyrus is involved in decision-making and social behavior. These findings are consistent with the established neuroscience literature and further support the clinical relevance of our model \cite{faris2024new, andreou2024caudate}. The saliency map provides insight into the neurobiological basis of schizophrenia, highlighting the importance of these key brain regions in the pathology of the disorder.

\begin{figure}
    \centering
    \includegraphics[width=0.9\linewidth]{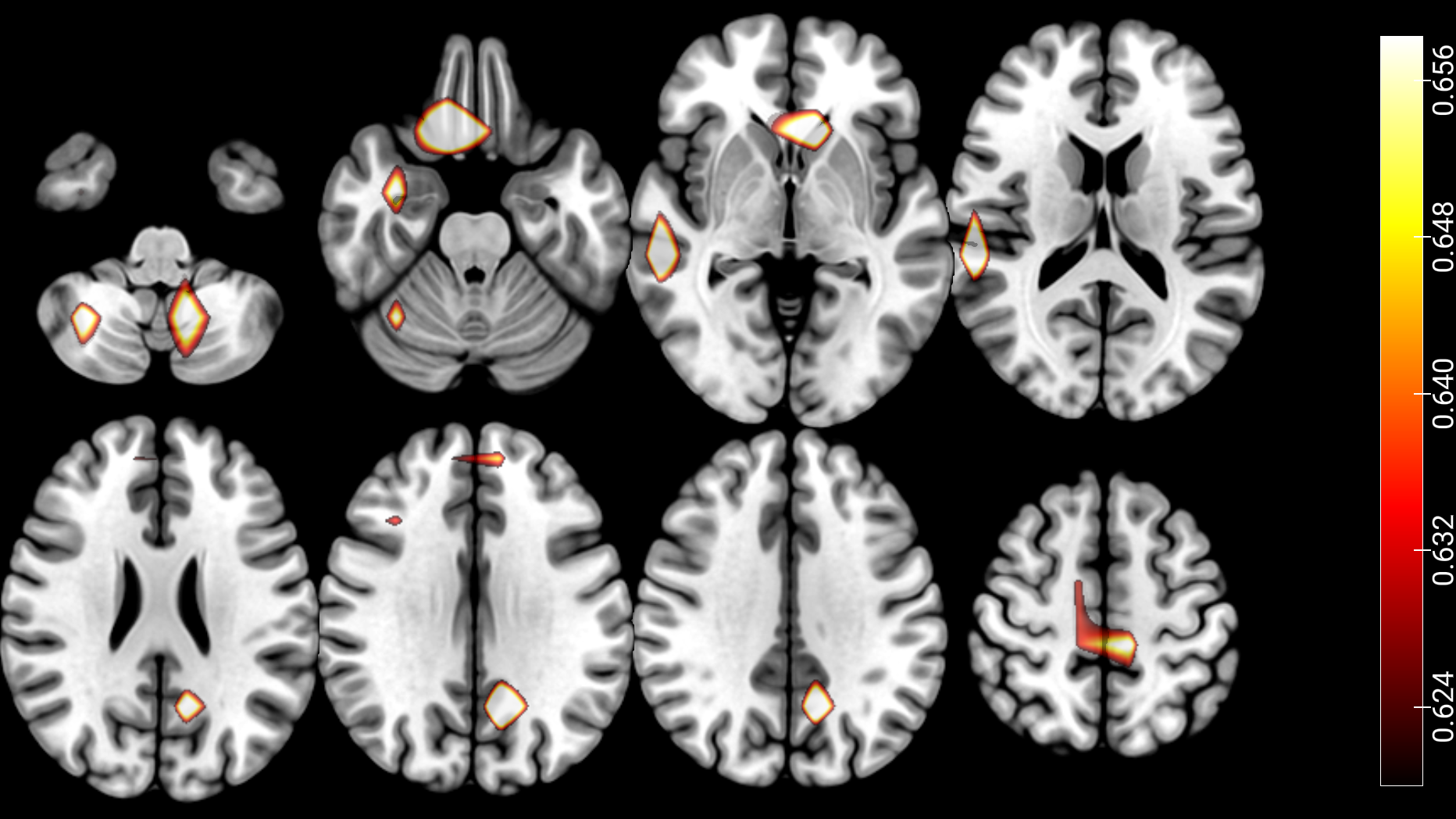}
    \caption{Saliency map showing key brain regions contributing to schizophrenia classification, including the cerebellum, caudate, precuneus, and superior frontal orbital gyrus, all associated with motor control, cognition, and emotional regulation.}
    \label{fig:enter-label}
\end{figure}

\section{Conclusions}

The MultiViT2 model offers a structured and transparent pipeline for predicting brain diseases and identifying biomarkers, providing a robust framework for deeper understanding of brain disorders. Its incorporation of LFFM enhances portability, addressing limited data availability in some studies, while the data augmentation module, particularly the pre-trained autoencoder, enables easy transfer and fine-tuning in 3D MRI environments for targeted optimization. This versatility makes the model adaptable to various medical imaging tasks. Future work will focus on applying MultiViT2 to a wider range of brain disorders, validating its generalizability and improving its robustness for medical diagnostics.

% References should be produced using the bibtex program from suitable
% BiBTeX files (here: strings, refs, manuals). The IEEEbib.bst bibliography
% style file from IEEE produces unsorted bibliography list.
% ------------------------------------------------------------------------- 
\bibliographystyle{IEEEbib}
\bibliography{strings,refs}

\end{document}